# The role of project science in the Chandra X-ray Observatory

Stephen L. O'Dell [a]* & Martin C. Weisskopf [a]

[a] NASA Marshall Space Flight Center, NSSTC/VP62, 320 Sparkman Dr., Huntsville, AL 35805

## ABSTRACT

The *Chandra X-ray Observatory*, one of NASA's Great Observatories, has an outstanding record of scientific and technical success. This success results from the efforts of a team comprising NASA, its contractors, the Smithsonian Astrophysical Observatory, the instrument groups, and other elements of the scientific community—including the thousands of scientists who utilize this powerful facility for astrophysical research. We discuss the role of NASA Project Science in the formulation, development, calibration, and operation of the *Chandra X-ray Observatory*. In addition to serving as an interface between the scientific community and the Project, Project Science performed what we term "science systems engineering". This activity encompasses translation of science requirements into technical requirements and assessment of the scientific impact of programmatic and technical trades. We briefly describe several examples of science systems engineering conducted by *Chandra* Project Science.

## 1. INTRODUCTION

On 1999 July 23 4:31 UTC, NASA's shuttle mission STS-93 launched (Figure 1) the *Chandra X-ray Observatory*. About 8 hours later, the orbiter *Columbia* deployed *Chandra* from its payload bay. Another hour later, the attached Inertial Upper Stage (IUS) fired and separated, sending *Chandra* toward a high elliptical orbit. On August 7, the fifth burn of *Chandra*'s integral propulsion system placed it into a 63.5-hour (80,800-km semi-major axis) orbit. Then, on August 12, the Observatory's forward door opened, exposing the focal plane to celestial x rays (Figure 2)—*first light!*

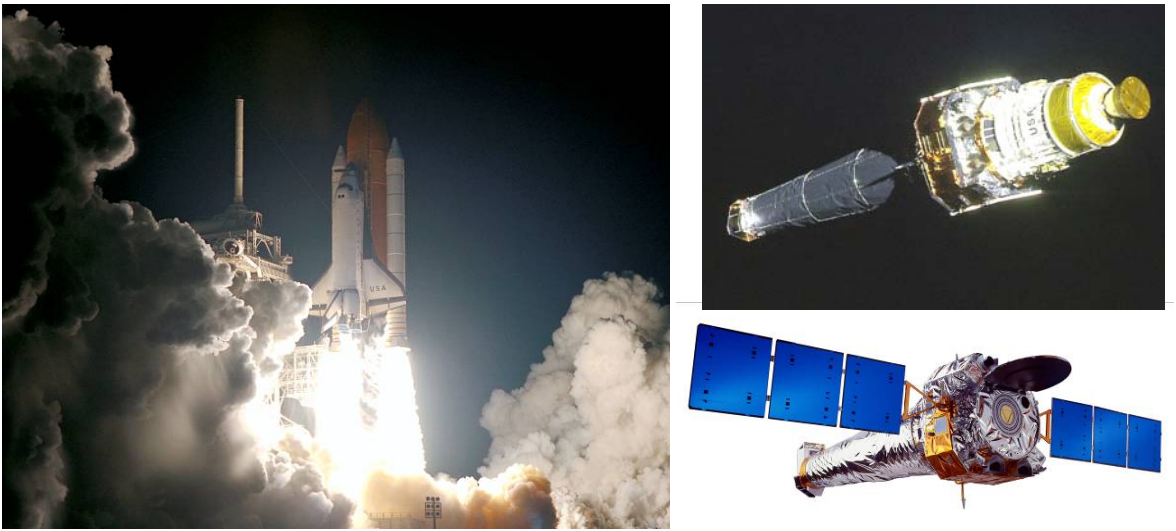

Figure 1: Start of orbital operations of the *Chandra X-ray Observatory*. Left photo (NASA) displays the launch of the STS-93 mission from the Kennedy Space Center (KSC). Upper right photo (NASA) shows the free spacecraft in stowed configuration with the IUS attached to the forward (optics) end of the Observatory). Lower right artist's rendering (NGST) illustrates the operational Observatory with solar arrays deployed and forward (sunshade) door open.

* Contact author SLO: Steve.O'Dell@nasa.gov; phone 1 256-961-7776; fax 1 256-961-7213; wwwastro.msfc.nasa.gov

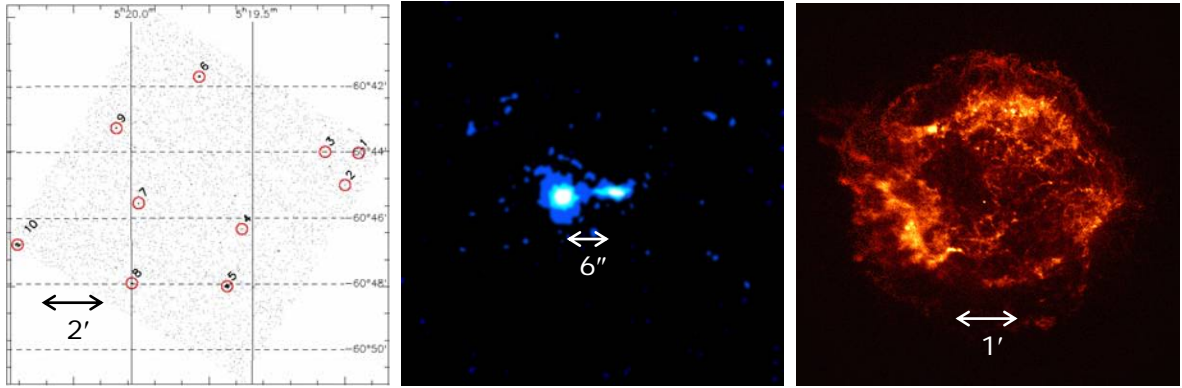

Figure 2: First light! Left x-ray event map (NASA/MSFC) displays *Chandra*'s first field[1] (with 10 detected sources—9 *Chandra*-discovered—on the central CCD), obtained upon opening the Observatory's forward door, before focusing and implementing fine-attitude control. Center smoothed x-ray event map (NASA/CXC) shows the quasar PKS 0637-752 (with *Chandra*-discovered x-ray jet[2]), used to focus the x-ray optics. Left x-ray image (NASA/CXC) of the Cassiopeia A supernova remnant (with *Chandra*-discovered central point source[3]) is the official first-light image.

Such a complex and magnificent achievement resulted from the dedicated effort of hundreds of individuals in dozens of organizations within NASA, scientific institutions, and industry. In this paper, we discuss the role of NASA project science in the *Chandra X-ray Observatory*. First (§2), we give background information on the Observatory, the Program, and Project Science. Next (§3), we describe project-science activities during each mission phase—formulation, development, calibration, and operations. Finally (§4), we present concluding remarks, highlighting some factors that we believe contributed to a highly successful, large scientific project.

## 2. BACKGROUND

The *Chandra X-ray Observatory*[4,5,6]—nee the Advanced X-ray Astrophysics Facility[7,8] (AXAF)—is the x-ray component of NASA's Great Observatories. Three of the four observatories—*Chandra*, *Hubble Space Telescope*, and *Spitzer Space Telescope*—are currently operational, providing astronomical observations at x-ray, at far-ultraviolet through near-infrared, and at infrared wavelengths, respectively. The fourth—the *Compton Gamma-Ray Observatory*—safely de-orbited on 2000 June 4. Here we provide background information on the *Chandra* Observatory (§2.1), the Program (§2.2), and Project Science (§2.3).

### 2.1. Chandra X-ray Observatory

The *Chandra* payload—Observatory, IUS, and deployment platform and support structure—aboard the *Columbia* orbiter was the most massive (22,750 kg = 50,160 lb) and longest (17.3 m = 57.0 ft) launched by a NASA space shuttle. About 60% (13,870 kg = 30,580 lb) of this payload mass was in the IUS, which provided the primary impulse to raise the Observatory to a highly elliptical orbit. Through the extensive use of lightweight carbon-fiber reinforced plastics, the Observatory itself is rather light (dry mass 4,800 kg = 10,560 lb) for its size (13.8 m = 45.3 ft long by 19.5 m = 64.0 ft wingspan), initially carrying about a ton (980 kg = 2,160 lb) of propellant, mainly for establishing the 63.5-h operational orbit (initially 10 Mm = 6,200 mi perigee altitude by 140 Mm = 86,900 mi apogee altitude, at 28.5° inclination). The *Chandra X-ray Observatory* (Figure 3) comprises three (3) major systems—the Spacecraft Module (§2.1.1), the Telescope System (§2.1.2), and the Integrated Science Instruments Module (ISIM, §2.1.3)

### 2.1.1. Spacecraft Module

The Spacecraft Module azimuthally wraps the mirror end of the Telescope System (§2.1.2). Seven subsystems constitute this system:

1. Pointing Control & Aspect Determination (PCAD) subsystem [attitude determination, solar-array control, slewing, pointing and dithering control, and momentum management]
2. Communication, Command, & Data Management (CCDM) subsystem [communications, command storage and processing, data acquisition and storage, computation support, timing reference, and primary-power switching]

3. Electrical Power Subsystem (EPS) [primary-power generation, regulation, storage, distribution, conditioning, and control]
4. Thermal Control Subsystem (TCS) [passive thermal control (where possible), heaters, and thermostats]
5. Structures and mechanical subsystem [spacecraft structures, mechanical interfaces amongst spacecraft subsystems and with the Telescope System]
6. Propulsion subsystem
    a. Integral Propulsion Subsystem (IPS) [disabled upon reaching final orbit]
    b. Momentum Unloading Propulsion Subsystem (MUPS) [maintain nominal reaction-wheel angular momentum]
7. Flight software subsystem [algorithms for attitude determination and control, command and telemetry processing and storage, and thermal and electrical power monitoring and control]

Attached to the outside of the Spacecraft Module is the Electron, Proton, and Helium Instrument[9] (EPHIN). This flight spare from the ESA–NASA Solar & Heliospheric Observatory (SOHO) serves as a radiation monitor[10] for the Observatory, as well as a space-physics science instrument[11,12].

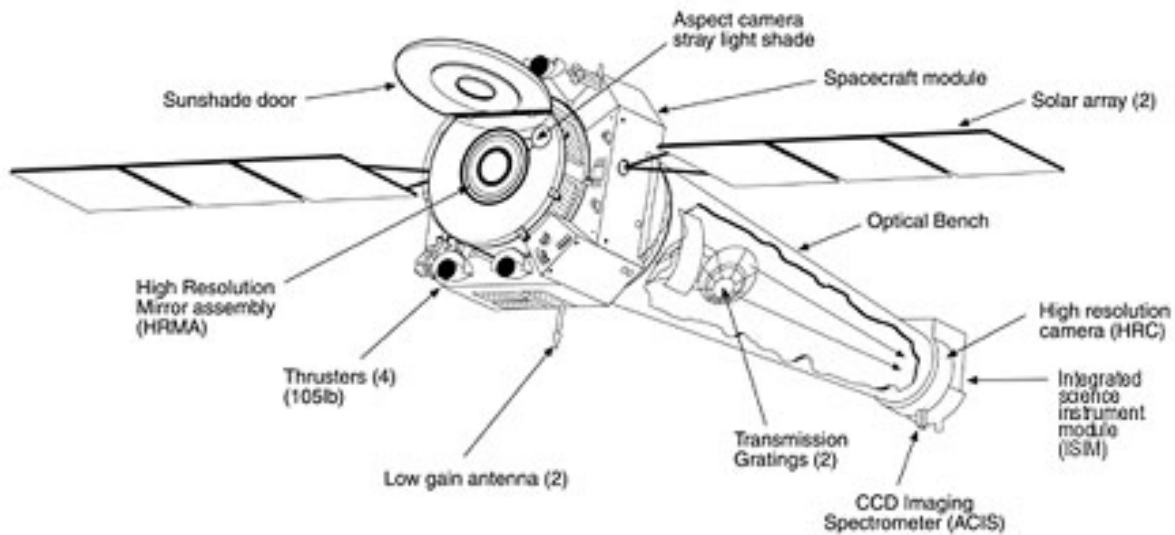

Figure 3: Schematic drawing (NGST) of the *Chandra X-ray Observatory*. Major systems are the Spacecraft Module, the Telescope System, and the Integrated Science Instruments Module (ISIM). Key components of the x-ray optical system are the HRMA, the two OTGs (LETG and HETG), and the two focal-plane instruments (HRC and ACIS).

### 2.1.2. Telescope System

The Telescope System empowers the Observatory with its unique capabilities for sub-arcsecond x-ray imaging and high-resolution dispersive spectroscopy. This system includes the following assemblies and subsystems:

1. High-Resolution Mirror Assembly[13,14] (HRMA) [4 iridium-coated[15,16] Wolter-1 (grazing-incidence parabolic–hyperbolic) mirror pairs, 10-m focal length, 0.6–1.2-m diameter; precision figured, polished, aligned[17], and assembled for sub-arcsecond imaging]
2. Aspect Camera Assembly[18] (ACA) [visible-light telescope and CCD camera]
3. Fiducial Transfer Subsystem (FTS) [mapping of x-ray focal plane onto visible-light ACA, for accurate ($< 0.3″$) post-factor aspect determination]
4. Telescope Forward Thermal Enclosure (TFTE) [maintenance of stable thermal environment[19] for the HRMA]
5. Optical Bench Assembly[20] (OBA) [stable, lightweight carbon-fiber-reinforced cyanate-ester metering structure]
6. Objective Transmission Grating (OTG) subsystem [insertion of either OTG (or none) into optical path]
    a. Low-Energy Transmission Grating[21] (LETG) [540 free-standing gold-bar micro-lithographic gratings, arranged on a Rowland toroid and optimized for performance below about 1 keV]
    b. High-Energy Transmission Grating[22,23] (HETG) [336 polyimide-supported micro-lithographic gratings (two types, differing dispersions), arranged on a Rowland toroid and optimized for performance above about 1 keV]

### 2.1.3. Integrated Science Instruments Module

The Integrated Science Instrument Module (ISIM) is located at the aft (detector) end of the Observatory. This system includes the following assemblies and subsystems:

1. Science Instrument Module[24] (SIM)
   a. Housing and mechanical interface to focal-plane detectors [thermal and stray-light isolation of detectors]
   b. Focus structure and mechanism [(axial) focus adjustment]
   c. Translation table and mechanism [transverse positioning and exchange of detector systems]
2. Focal-Plane Science Instruments (FPSI)
   a. High-Resolution Camera[25] (HRC) [microchannel-plate detectors—HRC-I[26] for high-speed high-resolution imaging and HRC-S[27] for reading x-ray spectra dispersed by an OTG (usually LETG)]
   b. Advanced CCD Imaging Spectrometer[28] (ACIS) [CCD arrays—2×2 ACIS-I for moderate-spectral-resolution high-resolution imaging and 1×6 ACIS-S for reading x-ray spectra dispersed by an OTG (usually HETG)]

### 2.2. Chandra Program

As is the case for most major scientific programs, the *Chandra*-AXAF program has a long history (§2.2.1). Likewise, its success results from the contributions of many organizations (§2.2.2).

### 2.2.1. Chronology

The *Chandra* program formally began with a 1976 unsolicited proposal to NASA, "The Study of the 1.2 Meter X-ray Telescope National Space Observatory". With Principal Investigator Riccardo Giacconi and Co-Investigator Harvey Tananbaum, the Smithsonian Astrophysical Observatory (SAO, Cambridge MA) submitted this proposal for a large high-resolution (sub-arcsecond) x-ray telescope to follow the second High-Energy Astrophysics Observatory (HEAO-2), then nearing completion. Soon thereafter, NASA Headquarters assigned project management to the Marshall Space Flight Center (MSFC, Huntsville AL) and named a pre-formulation Science Working Group (SWG) chaired by Dr. Giacconi. Nearly 25 years later, Dr. Giacconi, recognized as the "father of x-ray astronomy", shared the 2002 Nobel Prize in Physics for "pioneering contributions to astrophysics, which have led to the discovery of cosmic x-ray sources".

In 1978, HEAO-2—known as the *Einstein Observatory*—launched and operated for about 2.5 years. MSFC, SAO, and prime contractor TRW (Redondo Beach CA) had developed HEAO-2, the first focusing telescope for x-ray astronomy. Initially a Principal-Investigator (PI) mission—i.e., all data rights belonging to the instrument PIs—HEAO-2 became the first x-ray mission to offer a guest-observer program, establishing the *Einstein Observatory* as a model for future space-astronomy facilities. In 1981, the decadal survey[29] of the National Academy of Sciences named the Advanced X-ray Astrophysics Facility (AXAF) as the top priority for astronomy. Subsequent to this strong endorsement and the success of the *Einstein Observatory*, NASA in 1983 released an announcement of opportunity (AO) for AXAF science instruments and in 1985 selected a telescope scientist, the science-instrument teams, and interdisciplinary scientists.

Recognizing the x-ray mirrors as the mission-critical technology, MSFC with SAO mission support and telescope science oversaw development of the Technology Mirror Assembly (TMA; ⅔-scale version of the AXAF innermost mirror pair) at Hughes Danbury Optical Systems (HDOS; Danbury CT). MSFC and SAO x-ray tested the phase-1 TMA at MSFC in 1985 and (after additional figuring and polishing) the phase-2 TMA in 1989. The success of the TMA project gave the Project confidence to proceed with production at HDOS of the outermost flight mirror pairs. In 1991, MSFC and SAO x-ray tested these mirrors still uncoated, in a configuration termed the Verification Engineering Test Article (VETA), at MSFC's newly refurbished X-Ray Calibration Facility (XRCF). This crucial test demonstrated sub-arcsecond x-ray imaging, a congressionally mandated prerequisite to continuing the mission development.

Meanwhile, in 1988 NASA authorized a new start, enabling MSFC to select competitively the prime contractor—TRW (now Northrop-Grumman Space Technology, NGST). Cost-saving measures in 1989 resulted in downsizing the Bragg-Crystal Spectrometer (BCS), one of the original four (4) focal-plane instruments. In 1991, MSFC completed the primary *Chandra*-AXAF team with the competitive selection of SAO to operate the AXAF Science Center (now the *Chandra* X-ray Center). Also in 1991, the next decadal survey[30] re-affirmed AXAF as "the highest-priority large program".

In 1991–1992, the AXAF team engaged in an intensive restructuring activity directed by NASA Headquarters, to reduce the cost of the mission and thus help ensure congressional support. This restructuring split the AXAF program into two satellites—AXAF-I (imaging) and AXAF-S (spectroscopy). AXAF-I was similar to the original AXAF (re-designated "AXAF-O"), but deleted two (2) of six (6) mirror pairs and two (2) focal-plane instruments—the micro-calorimeter X-Ray Spectrometer (XRS) and the downsized BCS (since 1989 a back-up to the XRS). With these deletions and substantial light-weighting of the spacecraft, the original low-earth-orbit shuttle-serviceable AXAF-O became the high-elliptical-orbit (non-serviceable) AXAF-I, which is now the *Chandra X-ray Observatory*. The AXAF-S satellite was to have been a lower-cost lower-resolution x-ray telescope, matched to the capabilities of the XRS, its only science instrument. In 1993 Congress deleted funding for AXAF-S noting that it could possibly fly on a Japanese mission. Indeed, the XRS did eventually fly on the Japan–US Astro-E2 *Suzaku* mission, launched in 2005.

After 1994, Congress and NASA kept close to the planned funding profile. Preparations and execution of the extensive x-ray test and calibration of the Observatory at MSFC's X-Ray Calibration Facility (XRCF) occurred in 1996–1997. After XRCF activities and a delay in integration and testing, TRW completed the flight system in Redondo Beach and shipped AXAF to KSC aboard an Air Force C-5 Galaxy. Then, in 1999, *Columbia* launched AXAF, renamed *Chandra* in honor of Subrahmanyan Chandrasekhar—renowned astrophysicist and 1983 Nobel Laureate in Physics.

### 2.2.2. Contributing organizations

Many organizations have contributed to the success of the *Chandra X-ray Observatory*. Here, we mention some of the contributions of NASA centers (§2.2.2.1), scientific institutions (§2.2.2.2), and commercial concerns (§2.2.2.3).

### 2.2.2.1. NASA centers

The Marshall Space Flight Center (MSFC, Huntsville AL) manages the *Chandra* Program and provides budgetary, technical, and scientific oversight/insight. During development, MSFC managed directly the prime, mirror-fabrication, science-instrument, mission-support, science-center, and off-line-system software contracts. MSFC also performed certain hardware and software engineering tasks supporting development and testing. Perhaps the most significant engineering activity was refurbishing and operating the X-Ray Calibration Facility (XRCF).

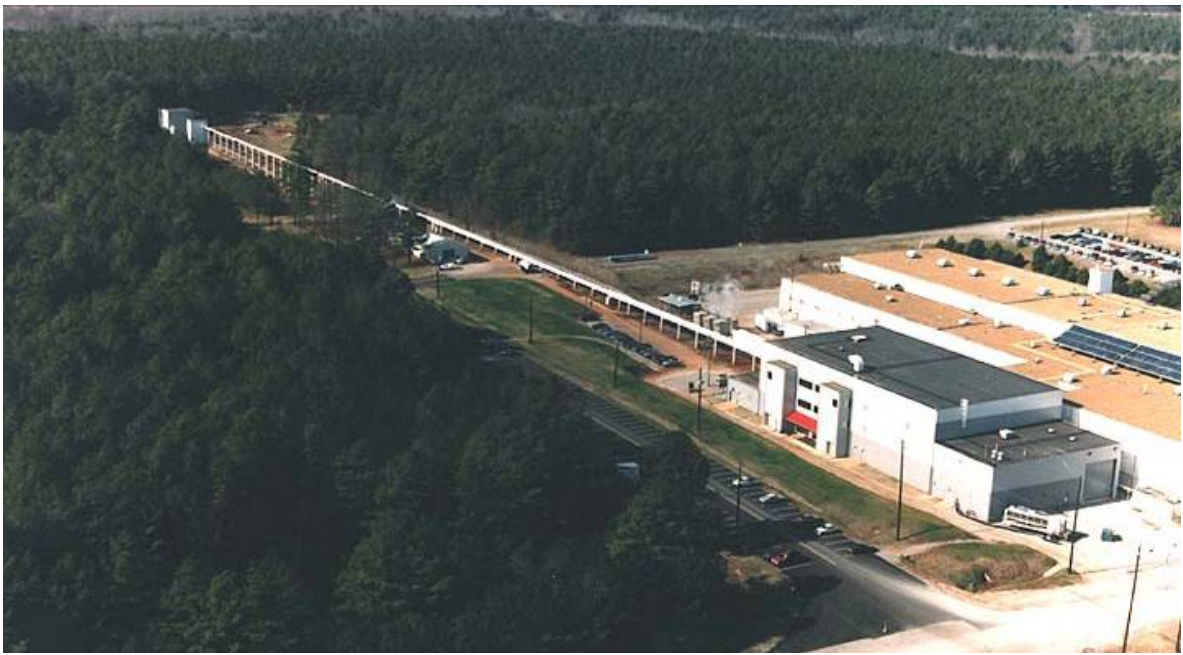

Figure 4: MSFC's X-Ray Calibration Facility (XRCF). The XRCF vacuum chamber (in lower right building) is large enough to hold any orbiter payload and sufficiently distant (530 m) from the sources (in upper left building) to produce the small source sizes ($< 0.2''$) needed to test and calibrate Chandra's sub-arcsecond x-ray optics.

The Goddard Space Flight Center (GSFC, Greenbelt MD) participated during mission formulation and early development as the PI institution for the X-Ray Spectrometer (XRS), one of the original focal-plane science instruments that was deleted following restructuring (§2.2.1). After launch, GSFC participated with MSFC and with the ACIS Team in studies addressing certain anomalies encountered during operations.

The Johnson Space Center (JSC, Houston TX) was responsible for STS-93 crew training and space-shuttle operations leading to the launch and deployment of the *Chandra X-ray Observatory*. The Kennedy Space Center (KSC, Cape Canaveral FL) integrated the *Chandra* payload (§2.1) into the orbiter *Columbia* and launched the space shuttle. The Jet Propulsion Laboratory (JPL, Pasadena CA) operates NASA's Deep-Space Network (DSN), used to communicate with the orbiting Observatory, approximately every 8 hours.

### 2.2.2.2. Scientific institutions

The Smithsonian Astrophysical Observatory (SAO; Cambridge MA) operates the *Chandra* X-ray Center (CXC), under contract to NASA MSFC. The CXC is directly responsible for science and flight operations, data systems, user support, and administration of the guest-observer program. During formulation and development, SAO provided scientific and mission support to MSFC under a previous contract. In addition, SAO provided the High-Resolution Camera (HRC; §2.1.3) and Telescope Science.

The Massachusetts Institute of Technology (MIT; Cambridge MA) partners with SAO in operating the CXC and also provided the High-Energy Transmission Grating (HETG; §2.1.3). With the PI institution Pennsylvania State University (PSU; University Park PA), MIT also developed the Advanced CCD Imaging Spectrometer (ACIS; §2.1.3). The Scientific Research Organization of the Netherlands (SRON; Utrecht, the Netherlands) provided the Low-Energy Transmission Grating (LETG; §2.1.3), with the Max-Planck-Institut für extraterrestrische Physik (MPE; Garching, Germany). Universität Kiel (Germany) supplied the Electron, Proton, & Helium Instrument (EPHIN; §2.1.1).

Many other scientific institutions also contributed through collaborations with the instrument teams and through Interdisciplinary Scientists (IDS) serving on the Science Working Group (SWG). Among these latter institutions are Cambridge University (Cambridge, UK), Johns Hopkins University (JHU; Baltimore MD), the University of Colorado (Boulder CO), and the University of Maryland (College Park MD), as well as NASA GSFC. Now, with the successful operation of the *Chandra X-ray Observatory*, scientists from hundreds of institution are contributing through their *Chandra* observations, data analysis, and research.

### 2.2.2.3. Commercial concerns

Northrop-Grumman Space Technology (NGST; Redondo Beach CA; formerly TRW) was the prime contractor for *Chandra*-AXAF and provided systems engineering, integration, and testing. In addition, NGST designed and built the Spacecraft Module (§2.1.1), the first of three major systems comprising the Observatory (§2.1). As prime contractor, NGST directly managed most major flight-system contracts. Now under contract to SAO, NGST provides the Flight Operations Team (FOT) at the CXC Operations Control Center (Cambridge MA) and sustaining engineering support.

ITT Space Systems (ITTSS; Rochester NY; formerly Eastman Kodak) was responsible for the Telescope System (§2.1.2), the second of three major Observatory systems (§2.1). Particularly challenging was alignment and assembly of the High-Resolution Mirror Assembly (HRMA) and development of the Optical Bench Assembly (OBA) using lightweight, dimensionally stable materials from Composite Optics Inc. (COI; San Diego CA; now a division of ATK ).

Ball Aerospace and Technologies Corp. (BATC; Boulder CO) was responsible for the Integrated Science Instrument Module (ISIM; §2.1.3), the third of three major Observatory systems (§2.1). In addition, BATC designed and built the (visible-light) aspect camera assembly (§2.1.2).

Hughes Danbury Optical Systems (HDOS; later Raytheon Optical Systems; now Goodrich Electro-Optical Systems) fabricated, figured, and polished Chandra's four precision x-ray grazing-incidence mirror pairs from Zerodur™ blanks from Schott Glaswerke (Mainz, Germany). Due to the need for early development of the critical mirror technology, MSFC managed the contract with HDOS and supplied the completed mirrors to the prime contractor (TRW; now NGST). TRW then contracted with Optical Coating Laboratory Inc. (OCLI; Santa Rosa CA; now JDS Uniphase Corp.) to coat the polished mirrors with sputtered iridium. After coating, Kodak integrated the coated mirrors into the HRMA.

Lockheed–Martin Aerospace (LMA; Denver CO; formerly Martin–Marietta) performed systems engineering and integration of the ACIS for MIT. Under contract to MSFC, Computer Sciences Corporation (CSC; El Segundo CA) developed the off-line software system for the CXC. With the cooperation of the US Air Force, MSFC procured the Inertial Upper Stage (IUS; §2.1 and Figure 1) from the Boeing Company (Chicago IL). Needless to say, many other commercial organizations—too numerous to mention—contributed to the mission's success.

## 2.3. Chandra Project Science

The role of project science in NASA programs has taken many forms, spanning a wide range in the degree of involvement. It its minimal form, project science is a lone individual, who makes an appearance at major reviews to present the scientific goals and accomplishments of the mission. In its more active form, project science is an integrated team that performs a variety of scientific functions in support of the project. Certainly, *Chandra* Project Science took the latter form, with a high degree of involvement through all phases of the mission (§3). Here, we present a brief history (§2.3.1) of *Chandra* Project Science and synopsize its roles (§2.3.2) in the mission.

### 2.3.1. History

As described earlier (§2.2.1), the AXAF (*Chandra X-ray Observatory*) program was the scientific successor to the successful HEAO-2 (*Einstein Observatory*) mission. The same organizations—MSFC and SAO—primarily responsible for HEAO-2 were also leading AXAF formulation. Consequently, the AXAF program profited from numerous "lessons learned" from the HEAO-2 mission. One of these lessons was that that "long-distance" project science was not effective: For the HEAO program, project management had been at one NASA center and project science at another.

The x-ray-astronomy community also felt that the program would benefit from scientific participation at all levels. In the scientists' view, such participation would not only ensure that scientific requirements were met, but could also facilitate and optimize scientific return without imposing unnecessarily stringent requirements that might significantly impact project resources—weight, power, volume, schedule, and cost. This philosophy was certainly controversial at the time. The NASA project culture was strongly engineering oriented: "Bring us your requirements and we will build it for you." The Soviet Space Science Program is said to have carried this approach to the extreme, wherein scientists interacted with projects only twice—once at the beginning and then upon delivery of the instruments. The many flaws of this approach include the assumption that initial scientific requirements are complete, require no iteration, and are inflexible. While it seems obvious that scientists, engineers, and managers must work together to optimize a scientific project within constraints, projects have not always achieved such teamwork. In almost all cases where it hasn't, the project has had problems—even if ultimately viewed as a success. Indeed, NASA HQ, MSFC, and SAO recognized that HEAO-2 had been such a project and sought to improve this relationship.

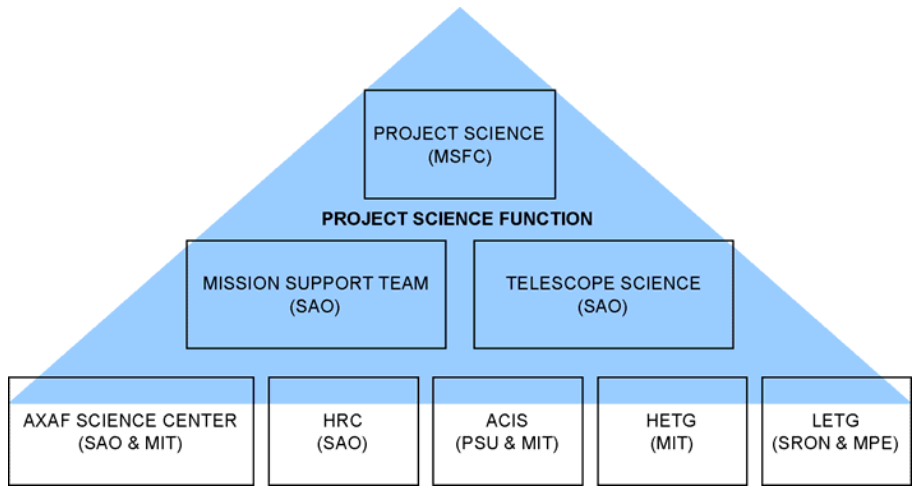

Figure 5: *Chandra* project-science function. During *Chandra*-AXAF mission development and calibration, MSFC Project Science, SAO Mission Support, and Telescope Science performed the expanded project-science function, working closely with the AXAF Science Center (*Chandra* X-ray Center) and science-instrument teams.

Consequently, the project-science function for AXAF differed substantially from that for HEAO-2 and also from that for the *Hubble Space Telescope* (HST), which was at a later stage of development. First, the AXAF Project Scientist was located at the same NASA Center—MSFC—as AXAF Project Management. Second, the Project Scientist (or delegate) was a member of each level-2 and level-3 control board, even those that did not obviously address science issues. Third, the project-science function resided, not in one or two individuals, but in an active team of scientists located at MSFC, at SAO, and at the PI institutions (Figure 5). Hence, upon initiating formulation of the mission in 1976–1977, MSFC enlisted an SAO Mission Support Team and the AXAF Project Scientist (MCW) arrived at MSFC. By 1980, the Project Scientist began building an MSFC project-science team that, at maximum staffing (during calibration at the XRCF), comprised 7 civil-servant and 3 contractor scientists. The 1985 selection of the permanent Science Working Group (SWG) included the Telescope Scientist (Leon Van Speybroeck, from SAO), who also shared in the project-science function with MSFC Project Science and SAO Mission Support. Also at this point, the Project Scientist, who had been serving as SWG Vice-Chair, became SWG Chair.

### 2.3.2. Roles

Project science serves as an interface between the science community and project. As such, we believe that project science is responsible both to the project and to the science community. Not only must project science represent science within the project, but it also should represent the project within the scientific community. This duality can become problematic, even confrontational, if the project elements fail to work together as a team toward a common objective. Our statement of the objective would be this: "Optimize the scientific performance of the mission within resource constraints and at an acceptable level of risk." Of course, even an enlightened project management might prefer to state the objective slightly differently: "Accomplish the mission within resource constraints and at an acceptable level of risk, while optimizing the scientific performance."

The overarching responsibility of project science is to ensure the scientific integrity of mission. For *Chandra*-AXAF Project Science, this meant doing more than the traditional activities of participating in the formulation of top-level science requirements, chairing the SWG, attending reviews, and making presentations. We believe that ensuring the scientific integrity requires active engagement at all levels. Thus, in addition to the traditional activities, *Chandra* Project Science performed what we term "science systems engineering".

Science systems engineering encompasses investigating science-performance trades, translating science requirements into technical specifications, monitoring the flow-down and implementation, and verifying that science requirements are satisfied. These activities involve identification of relevant issues, modeling and analysis, and testing. Science systems engineering (at both MSFC and SAO) played a significant role in the *Chandra* Project. It helped not only to avoid "requirements creep" but also to prevent erroneous flow-down of requirements to technical specifications that would be excessively costly, unnecessary, or (in a few interesting cases) detrimental to scientific performance.

A particularly instructional example concerned AXAF pointing requirements. Many members of MSFC project management and engineering had been active in development of the HST, for which precise and stable pointing was a major driver. Delivering a high-precision attitude-control system for HST was justifiably a matter of great pride. Because AXAF was also to achieve sub-arcsecond resolution, project engineering understandably strove for pointing precision and stability comparable to that of HST. However, the scientific requirements for AXAF pointing—$30''$ placement, $1''$ location, and $0.05''$ 10-s stability—are much less demanding than those for HST: *Chandra* x-ray images result from post-facto reconstruction of many individual short exposures, using a visible-light aspect camera (§2.1.1) to map x-ray focal-plane events onto the sky. For AXAF, the precise ultra-stable pointing of HST would have been not just unnecessary but detrimental. First, an ACIS (§2.1.3) x-ray CCD pixel is a non-trivial fraction of the telescope's point spread function (PSF), thus ultra-stable pointing would require calibrating individually each of 10-million ACIS pixels at a sub-pixel level. Second, if an observation had an x-ray source focused onto a gap between CCDs, that source would be missed. Finally, HRC (§2.1.3) microchannel plates are subject to charge depletion, whereby the gain of an individual pixel eventually drops if its total x-ray dose becomes large; hence, in effect, ultra-stable pointing would "burn a hole" into the detector. For these reasons, the *Chandra* Observatory operates in a "dither" mode, the pointing direction systematically executing a Lissajou pattern over an area comprising more than a thousand pixels.

### 3. MSFC PROJECT SCIENCE ACTIVITIES

MSFC Project Science, SAO Mission Support Team, and SAO Telescope Science each contributed actively to performing the *Chandra* project-science function. Here we specifically highlight activities of *Chandra* Project Science at MSFC during mission formulation (§3.1), development (§3.2), calibration (§3.3), and operations (§3.4). These activities encompassed both (more traditional) oversight/insight responsibilities and science-systems-engineering tasks. Space does not permit detailed description of any of these science-systems-engineering tasks; however, we have documented several of these studies in previous papers, as referenced below.

### 3.1. Formulation

During mission formulation, science oversight/insight included leading the definition of science requirements, vice-chairing the temporary SWG, and serving on the source evaluation board (SEB) for selection of the prime contractor. A particularly intense activity was leading the science team during restructuring of the AXAF program (§2.2.1).

Science-systems-engineering activities included explaining the relaxed pointing-stability requirement for AXAF aspect determination (§2.3.2), developing an independent ray-trace code for the AXAF telescope, and commencing several studies that eventually led to detailed scientific requirements and flow-down to technical specifications. During the 1991–1992 restructuring, MSFC Project Science performed modeling and analyses to support several mission trades— e.g., effective area for the various de-scoped mirror configurations for AXAF-I and the effective-area trade for iridium, gold, nickel coatings[31]. Remarkably, the higher observing efficiency of a highly elliptical orbit and the greater reflectance of iridium at the higher x-ray energies essentially compensated for the loss of 2 of the 6 mirror pairs, resulting in an AXAF Observatory that promised to be as scientifically productive as the original AXAF, with gold-coated optics and in a low-earth orbit. After restructuring, MSFC Project Science led formulation of scientific requirements for the AXAF-S mission, developed an optical design for the AXAF-S telescope matched to XRS capabilities, and tested a prototype mirror—an electroformed nickel replica developed at MSFC.

### 3.2. Development

During mission development, science oversight/insight included chairing the permanent SWG and participating in technical reviews and interchange meetings. In addition, the Project Scientist or delegate sat on each level-2 and level-3 control board and served on the source evaluation board (SEB) for selection of the AXAF science center.

Science-systems-engineering activities included numerous modeling and analysis studies leading to detailed scientific requirements and flow down to technical specifications. In performing these activities MSFC Project Science worked closely both with scientists from the science-instrument, SAO Mission Support, and Telescope Science teams and with engineers at MSFC and at TRW. Several of the studies addressed requirements on particulate contamination[32,33] and on molecular contamination[34,35], which were particularly stringent to ensure ≈1% calibration and ground-to-orbit flux-scale transfer. Grazing-incidence optics are much more sensitive to particulate contamination than are normal-incidence optics: The small (0.5°–1° for *Chandra*) grazing angles result in grains shadowing a mirror surface area many times (of order 100 for *Chandra*) their own cross-sectional area; hence, the AXAF requirement on dust was about 2 orders of magnitude more severe than the HST requirement. Grazing-incidence x-ray reflection is much more sensitive to molecular contamination than is x-ray transmission. Consequently, AXAF requirements on molecular contamination on the mirrors are very stringent: Molecular contamination on the mirrors can change the HRMA's effective area by as much a few percent per nanometer of film thickness, at some energies (near atomic edges of the coating).

Many other studies involved working with the science-instrument teams to identify potential sources of background, followed by modeling and analysis to determine technical specifications that would adequately suppress those sources of background. Specific issues included shielding against stray x radiation, baffling stray infrared–ultraviolet light, filtering focused light, shielding against penetrating (hard-x-ray and proton[36]) radiation, and a magnetic broom for sweeping focused electrons. Project Science identified materials for shielding against penetrating radiation and worked closely with TRW systems engineers in strategically locating the shielding within the Observatory, which was inherently rather x-ray transparent due to its extensive use of lightweight (low-atomic-number) composites.

In an unusual role, MSFC Project Science also delivered some small flight-hardware components for the *Chandra X-ray Observatory*—the Flight Contamination Monitor (FCM). From our studies of the effects of contamination on x-ray reflectance[34,35], we identified a need to verify the transfer of the flux scale from ground calibration to on-orbit performance. Consequently, MSFC Project Science conceived and specified the FCM[37], a system of 16 radioactive calibration sources to be located on the forward contamination cover (FCC) and measured with the ACIS just prior to opening (permanently) the FCC/sunshade door. By virtue of its experience with radioactive calibration sources in developing x-ray instruments, the MSFC (Project Science) x-ray astronomers designed, specified, encapsulated, packaged, calibrated, and delivered the packaged radioactive sources—$^{109}$Cd (Ag-L$\alpha$ line at 3.0 keV) and $^{55}$Fe (Mn-K$\alpha$ line at 5.9 keV) e-capture sources—prepared by Isotope Products Laboratories (IPL; Valencia CA; now a division of Eckert & Ziegler AG, Berlin, Germany). In order to analyze focal-plane spectrometric images of the calibration sources, MSFC Project Science developed simulations using its HRMA ray-trace code. Finally, Project Science analyzed both the XRCF calibration data[38] and the flight data[39,40] to verify flux-scale transfer at better than 2%.

### 3.3. Calibration

MSFC Project Science had primary scientific responsibility for calibration and testing of at the XRCF[41,42] (§§2.2.1 and 2.2.2.1). This activity also served to verify the contractual performance specifications for the Observatory (Figure 6). Hence, Program and Project Management, prime contractor, and principal subcontractors recognized its importance—as, of course, did the AXAF scientific organizations. The Program had always planned extensive x-ray calibration and testing: HST's misfortune with its untested primary mirror ensured that no one seriously considered deleting or substantially reducing the x-ray test program.

In defining scientific requirements for x-ray calibration, Project Science led the Calibration Task Team (CTT) and wrote the calibration requirements document (AXAF 2229), which identified types of tests and requisite x-ray equipment—sources, detectors, etc. With TRW (NGST), Project Science co-led the Calibration Implementation Team (CIT) in planning and executing the calibration, culminating in the 6-month 24/7 effort at the XRCF that involved nearly every project element engaged in developing AXAF. In executing the 24/7 calibration test sequences, three MSFC Project-Science staff (Project Scientist, Deputy, and Calibration Scientist), the SAO Telescope Scientist, and the (SAO) Science Center Director rotated responsibility as shift lead scientist. After calibration, MSFC Project Science coordinated the Calibration Analysis Team (CAT), in resolving issues in the analysis of the calibration data. However, the AXAF Science Center (*Chandra* X-ray Center) had primary responsibility for the analysis of calibration data.

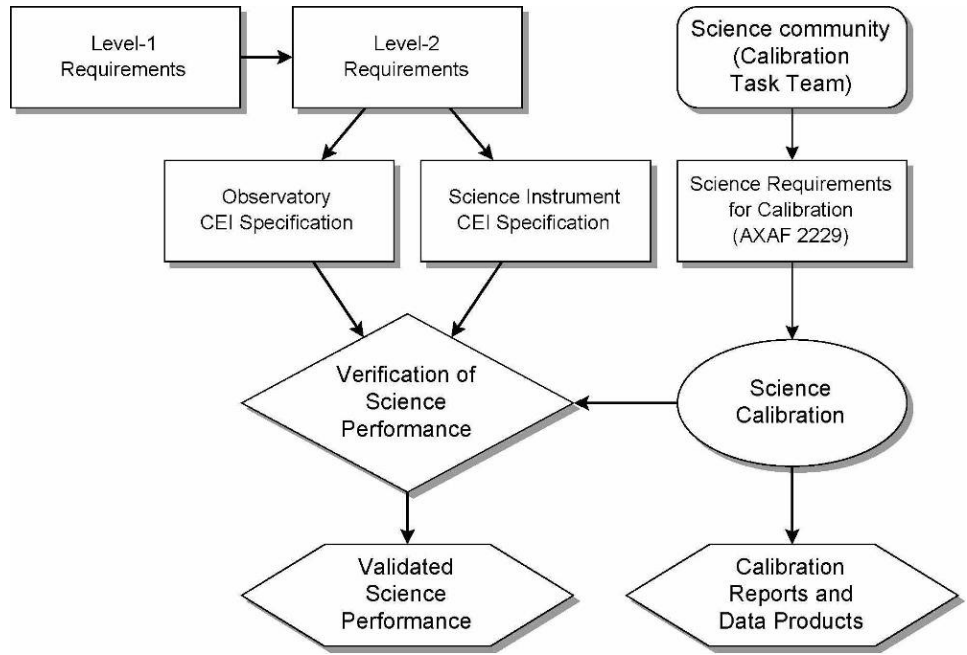

Figure 6: Requirements for x-ray calibration and testing. The x-ray testing at MSFC's XRCF served both as an x-ray calibration of the Observatory and as a verification of scientific-performance requirements in the contract end item (CEI) specifications for the Observatory and for the science instruments.

During the years preceding the calibration at the XRCF, SAO developed and characterized the (ground) X-ray Detector System (XDS), MSFC engineering developed the X-ray Source System (XSS) per AXAF2229, and Project Science characterized the XSS. The extensive XSS characterization encompassed x-ray measurements and data analysis for potential operating parameters of all the x-ray sources:

1. Electron-impact point source (EIPS) and Filter-Wheel Assembly, developed by MSFC;
2. Double-crystal monochromator (DCM), developed by the National Institute of Standards and Technology (NIST), with a high-power Rigaku rotating-anode source (RAS);
3. High-Resolution Erect-Field Spectrometer (HiREFS) used as a grating monochromator, purchased from Hettrick Scientific, with a high-power Rigaku rotating-anode source (RAS); and
4. Penning ionized-gas source (PIGS), provided by SAO.

Complementing the EIPS characterization, Project Science developed a software model to simulate x-ray spectra from the electron-impact point source and produced a spectral catalog for 21 anode targets and matched filters. As a tool for planning the calibration time-line, Project Science developed count-rate predictors based upon fluxes determined from characterization and simulation of the XSS, detector efficiencies provided by developers or other experts, HRMA effective area from ray-trace simulations by Telescope Science and by Project Science, science-instrument (SI) responses from the SI teams, and X-ray Detection System (XDS) responses from the SAO Mission Support Team. As an aid in executing the calibration tests, Project Science produced a software tool for displaying test configurations—X-ray source, EIPS target or monochromator wavelength, and filter; HRMA sub-aperture, grating (if any), off-axis angle, and de-focus; focal-plane SI imager or XDS and pinhole aperture, and scan. Finally, MSFC Project Science proposed and devised procedures for many calibration tests and conducted independent analyses to verify performance.

### 3.4. Operation

Now, during the operations phase, MSFC Project Science advises the MSFC Program Manager and the CXC Director (e.g., in assigning Directors Discretionary Time), serves as an ex officio member Users Group, and reviews CXC-generated documents—including the NASA Research Announcement (NRA) and the Proposers' Observatory Guide (POG). In addition, Project Science works with the CXC in establishing the integrity of the calibration of the scientific performance and in identifying and addressing anomalies or other issues.

Science systems engineering tasks during operations have focused primarily on anomalies in scientific performance of the instruments. The most significant such anomaly was rapid degradation, during initial operations, in the charge-transfer inefficiency (CTI) of 8 of the 10 ACIS CCDs—namely, those CCDs that are front-illuminated (FI). With the ACIS Team, MIT Lincoln Laboratory (Lexington MA), and the Air Force Research Laboratory (AFRL; Hanscom AFB MA), we helped establish the cause—radiation damage by low-energy (0.1-MeV) suprathermal protons Rutherford scattered off the x-ray mirrors onto the focal plane. Project Science led a detailed study of the 0.1-Mev proton environment for Chandra's orbit[43] and developed a numerical model simulating HRMA's transmission for such protons[44]. Fortunately, the Chandra Team rapidly formulated a radiation-management program[45,46]—most importantly, hiding the ACIS during radiation-belt transits—that reduced the rate of CTI degradation of the FI CCDs to acceptable levels (currently about 2% per year). To support the radiation-management program, Project Science and the MSFC Space Environments Group has developed the *Chandra* Radiation Model[47] (CRM), which estimates the low-energy-proton flux throughout *Chandra*'s orbit—magnetosphere, magnetosheath, and solar-wind regions—and can be driven by real-time space-weather data from the National Oceanic & Atmospheric Administration (NOAA) Space Environment Center (SEC; Bolder CO).

The second most significant anomaly in scientific performance is a loss in low-energy efficiency of all ACIS CCDs. With CXC and ACIS teams, we helped establish the cause—accumulation of molecular contamination onto the ACIS optical blocking filters (OBF), which operate at cold temperatures ($\approx$-55°C) that retard vaporization. Despite stringent contamination controls, the film is 30 times thicker than specified, yet totals < 0.5 g in the 5-tonne Observatory! Project Science supported an extensive CXC-led investigation[48] of this anomaly and potential approaches for dealing with it—including baking the ACIS instrument with on-board heaters designed for this purpose. Based upon previous experience and extensive ground testing, the ACIS Team and NGST concluded that a bake would be safe for the ACIS. In order to assess the effectiveness of a potential bake out, Project Science developed and utilized software to simulate molecular transport within the Observatory[49], using a geometrical model developed by NGST (for thermal analysis) and temperatures provided by NGST and LMA (ACIS systems integrator). Because the identity of the primary contaminant remains unknown[50] and temperatures of some critical surfaces are uncertain, we had to simulate contamination migration for a range of volatilities and surface temperatures. These simulations indicate a small parameter (volatility and temperature) space for a successful bake out—i.e., one that would clean the contaminated OBF. More importantly, however, they show a small but finite parameter space for leaving the OBF more contaminated than before the bake: This follows from an inability to make the OBF the warmest surface during the bake. The possibility of worsening the problem contributed significantly to the decision of the CXC and Chandra Program not to bake the ACIS at this time.

The third scientific-performance anomaly is an HRC timing error that, after a correction algorithm from the HRC Team, impacts few observations. MSFC Project Science independently discovered the symptoms of this problem in its observation[51] of the Crab Pulsar and devised an interim data-handling procedure for mitigating its effect. The HRC Team rapidly identified the cause and the correction algorithm. Although the impact of this anomaly is relatively minor, it demonstrates a benefit of another important project-science activity—namely, conducting scientific research.

Through conducting scientific research—especially *Chandra* observations—Project Science gains a users' perspective of the Observatory. Having both a user's view and being familiar with most aspects of the hardware and software, Project Science is in a position to identify issues impacting the general investigator. We note that MSFC Project Science has no guaranteed *Chandra* observing time, as do the instruments teams and the interdisciplinary scientists: Project Science competes for peer-reviewed observing time, as do CXC scientists and general investigators.

## 4. CONCLUDING REMARKS

We conclude with a few observations of some factors that we believe contributed to the success of this large scientific project. First and foremost, Project Management encouraged teamwork amongst all elements of the project and amongst all project cultures—managers, engineers, and scientists. In our view, effective teamwork requires a combination of a shared goal and communication, professional competency and dedication, mutual respect and cooperation, and engagement and constructive criticism. Overall, managers, engineers, and scientists from all organizations worked together well toward the Project's objectives.

From our perspective, Project Management's recognition of the need for an expanded project-science function was also critical. Distributing this function amongst three groups—MSFC Project Science, SAO Mission Science, and Telescope Science—was valuable, in that it ensured a direct conduit to MSFC Project Management, assembled requisite (often pre-existing) expertise, and provided cross-checking and redundancy. The synergy and checks and balances between MSFC Project Science and SAO Mission Support provided benefits greater than the sum of the parts. Monolithic groups can become parochial, resulting in overconfidence and uncritical acceptance of colleagues' work: As remarked above, although cooperation is valuable, so too is constructive criticism. To quote Ronald Reagan, "Trust, but verify."

Finally, Management involved Project Science in all aspects of the Program, not just those that were manifestly "science". Not only did this contribute to an environment fostering teamwork, but it brought to Project Science visibility into issues that otherwise might have escaped attention. Thus, through a project-science team effort and active engagement in the Project, basic requirements remained stable throughout the program (more than 20 years) and the Program successfully realized all identified scientific requirements.

## ACKNOWLEDGEMENTS

We gladly recognize the other past and present members of MSFC Project Science—Drs. Robert A. Austin, Charles R. Bower, Roger W. Bussard, Ronald F. Elsner, Marshall K. Joy, Jeffery J. Kolodziejczak, Brian D. Ramsey, Martin E. Sulkanen, Douglas A. Swartz, Allyn F. Tennant, Alton C. Williams, & Galen X. Zirnstein, and Mr. Darell Engelhaupt. We gratefully acknowledge the many contributions to the project-science function of the SAO Mission Support Team (led successively by Drs. Riccardo Giacconi, Harvey D. Tananbaum, & Daniel A. Schwartz) and Telescope Science (led by the late Dr. Leon Van Speybroeck, our dear friend and colleague). We appreciate efforts of Project Management in building the *Chandra* Team. Finally, we thank everyone who contributed to the success of *Chandra X-ray Observatory*.## REFERENCES

[1] M. C. Weisskopf, T. L. Aldcroft, R. A. Cameron, P. Gandhi, C. Foellmi, R. F. Elsner, S. K. Patel, K. Wu, & S. L. O'Dell, "The First Chandra Field", Astrophys. J., **637**, 682–692, 2006.

[2] D. A. Schwartz,, H. L. Marshall, J. E. J. Lovell, B. G. Piner, S. J. Tingay, M. Birkinshaw, G. Chartas, M. Elvis, E. D. Feigelson, K. K. Ghosh, D. E. Harris, H. Hirabayashi, E. J. Hooper, D. L. Jauncey, K. M. Lanzetta, S. Mathur, R. A. Preston, W. H. Tucker, S. Virani, B. Wilkes, & D. M. Worrall, "Chandra Discovery of a 100-kiloparsec X-ray Jet in PKS 0637-752", Astrophys. J., **540**, L69–72, 2000.

[3] H. Tananbaum et al., "Cassiopeia A", IAU Circ., **7246**, 1, 1999.

[4] M. C. Weisskopf, H. D. Tananbaum, L. P. Van Speybroeck, & S. L. O'Dell, "Chandra X-ray Observatory (CXO): overview", Proc. SPIE, **4012**, 2–16, 2000.

[5] M. C. Weisskopf, B. Brinkman, C. Canizares, G. Garmire, S. Murray, & L. P. Van Speybroeck, "An Overview of the Performance and Scientific Results from the Chandra X-ray Observatory", Pub. Astron. Soc. Pacific, **114**, 1–24, 2002.

[6] M. C. Weisskopf, "The Chandra X-ray Observatory: An overview", Adv. Space Research, **32**, 2005–2011, 2003.Proc. SPIE 6271-07-12

[7] M. C. Weisskopf, S. L. O'Dell, R. F. Elsner, & L. P. Van Speybroeck, "Advanced X-ray Astrophysics Facility (AXAF): an overview", Proc. SPIE, **2515**, 312–329, 1995.

[8] Weisskopf, M. C., S. L. O'Dell, & L. P. Van Speybroeck, "Advanced X-ray Astrophysics Facility (AXAF)", Proc. SPIE, **2805**, 2–7, 1996.

[9] R. Muller-Mellin, H. Kunow, V. Fleissner, E. Pehlke, E. Rode, N. Roschmann, C. Scharmberg, H. Sierks, P. Rusznyak, S. McKenna-Lawlor, I. Elendt, J. Sequeiros, D. Meziat, S. Sanchez, J. Medina, L. del Peral, M. Witte, R. Marsden, & J. Henrion, "COSTEP—Comprehensive Suprathermal and Energetic Particle Analyser", Solar Phys., **162**, 483–504, 1995.

[10] S. N. Virani, R. Mueller-Mellin, P. P. Plucinsky, & Y. M. Butt, "Chandra X-ray Observatory's radiation environment and the AP-8/AE-8 model", Proc. SPIE, **4012**, 669–680, 2000.

[11] R. Mueller-Mellin, J. B. Blake, & D. Baker, "Coordinated Energetic Particle Measurements Using Chandra, Cluster, and Polar", AGU Fall Meeting Abs., 2003.

[12] J. B. Blake, R. Mueller-Mellin, J. A. Davies, X. Li, & D. N. Baker, "Global observations of energetic electrons around the time of a substorm on 27 August 2001", J. Geophys. Res. (Space Phys.), **110**, A06214, 2005.

[13] T. J. Gaetz, W. A. Podgorski, L. M. Cohen, M. D. Freeman, R. J. Edgar, D. Jerius, L. P. Van Speybroeck, P. Zhao, J. J. Kolodziejczak, & M. C. Weisskopf, "Focus and alignment of the AXAF optics", Proc. SPIE, **3113**, 77–88, 1997.

[14] L. P. Van Speybroeck, D. Jerius, R. J. Edgar, T. J. Gaetz, P. Zhao, & P. B. Reid, "Performance expectation versus reality", Proc. SPIE, **3113**, 89–104, 1997.

[15] J. S. Bessey & J. A. Roth, "Sputtered iridium coatings for grazing-incidence x-ray reflectance", Proc. SPIE, **2011**, 12–17, 1994.

[16] R. J. Bruni, A. M. Clark, J. F. Moran, D. T. Nguyen, S. E. Romaine, D. A. Schwartz, & L. P. Van Speybroeck, "Verification of the coating performance for the AXAF flight optics based on reflectivity measurements of coated witness samples", Proc. SPIE, **2805**, 301–310, 1996.

[17] M. Waldman, "Alignment test system for AXAF-I's high-resolution mirror assembly", Proc. SPIE, **2515**, 330–339, 1995.

[18] D. Morris, T. L. Aldcroft, R. A. Cameron, M. L. Cresitello-Dittmar, & M. Karovska, "Analysis of the Chandra x-ray observatory aspect camera PSF and its application to post-facto pointing aspect determination", Proc. SPIE, **4477**, 254–264, 2001.

[19] K. A. Havey, G. L. Compagna, & N. Lynch, "Precision thermal control trades for telescope systems", Proc. SPIE, **3356**, 1127–1138, 1998.

[20] C. R. Olds & R. P. Reese, "Composite structures for the Advanced X-ray Astrophysics Facility (AXAF) Telescope", Proc. SPIE, **3356**, 910–921, 1998.

[21] B. C. Brinkman, T. Gunsing, J. S. Kaastra, R. van der Meer, R. Mewe, F. B. Paerels, T. Raassen, J. van Rooijen, H. W. Braeuninger, V. Burwitz, G. D. Hartner, G. Kettenring, P. Predehl, J. J. Drake, C. O. Johnson, A. T. Kenter, R. P. Kraft, S. S. Murray, P. W. Ratzlaff, & B. J. Wargelin, "Description and performance of the low-energy transmission grating spectrometer on board Chandra", Proc. SPIE, **4012**, 81–90, 2000.

[22] T. H. Markert, C. R. Canizares, D. Dewey, M. McGuirk, C. S. Pak, & M. L. Schattenburg, "High-Energy Transmission Grating Spectrometer for the Advanced X-ray Astrophysics Facility (AXAF)", Proc. SPIE, **2280**, 168–180, 1994.

[23] C. R. Canizares, J. E. Davis, D. Dewey, K. A. Flanagan, E. B. Galton, D. P. Huenemoerder, K. Ishibashi, T. H. Markert, H. L. Marshall, M. McGuirk, M. L. Schattenburg, N. S. Schulz, H. I. Smith, & M. Wise, "The Chandra High-Energy Transmission Grating: Design, Fabrication, Ground Calibration, and 5 Years in Flight", Pub. Astron. Soc. Pacific, **117**, 1144–1171, 2005.

[24] M. A. Skinner & S. P. Jordan, "AXAF: the Science Instrument Module", Proc. SPIE, **3114**, 2–10, 1997.

[25] S. S. Murray, G. K. Austin, J. H. Chappell, J. J. Gomes, A. T. Kenter, R. P. Kraft, G. R. Meehan, M. V. Zombeck, G. W. Fraser, & S. Serio, "In-flight performance of the Chandra high-resolution camera", Proc. SPIE, **4012**, 68–80, 2000.

[26] A. T. Kenter, J. H. Chappell, R. P. Kraft, G. R. Meehan, S. S. Murray, M. V. Zombeck, K. T. Hole, M. Juda, R. H. Donnelly, D. Patnaude, D. O. Pease, C. Wilton, P. Zhao, G. K. Austin, G. W. Fraser, J. F. Pearson, J. E. Lees, A. N. Brunton, M. Barbera, A. Collura, & S. Serio, "In-flight performance and calibration of the Chandra high-resolution camera imager (HRC-I)", Proc. SPIE, **4012**, 467–492, 2000.

[27] R. P. Kraft, J. H. Chappell, A. T. Kenter, G. R. Meehan, S. S. Murray, M. V. Zombeck, R. H. Donnelly, J. J. Drake, C. O. Johnson, M. Juda, D. Patnaude, D. O. Pease, P. W. Ratzlaff, B. J. Wargelin, P. Zhao, G. K. Austin, G. W. Fraser, J. F. Pearson, J. E. Lees, A. N. Brunton, M. Barbera, A. Collura, & S. Serio, "In-flight performance and calibration of the Chandra high-resolution camera spectroscopic readout (HRC-S)", Proc. SPIE, **4012**, 493–517, 2000.

[28] G. P. Garmire, M. W. Bautz, P. G. Ford, J. A. Nousek, & G. R. Ricker Jr., "Advanced CCD imaging spectrometer (ACIS) instrument on the Chandra X-ray Observatory", Proc. SPIE, **4851**, 28–44, 2003.

[29] Astronomy Survey Committee, *Astronomy and Astrophysics for the 1980's. Volume 1 — Report of the Astronomy Survey Committee*, National Academy Press, Washington DC, 1982.

[30] Astronomy and Astrophysics Survey Committee, *The Decade of Discovery in Astronomy and Astrophysics*, National Academy Press, Washington DC, 1991.

[31] R. F. Elsner, S. L. O'Dell, & M. C. Weisskopf, "Effective area of the AXAF X-ray telescope — Dependence upon dielectric constants of coating materials", J. X-Ray Sci. & Tech., **3**, 35–44, 1991.

[32] J. J. Kolodziejczak, S. L. O'Dell, R. F. Elsner, & M. C. Weisskopf, "Evidence for dust contamination on the VETA-1 mirror surface", Proc. SPIE, **1742**, 162–170, 1993.

[33] S. L. O'Dell, R. F. Elsner, J. J. Kolodziejczak, M. C. Weisskopf, J. P. Hughes, & L. P. Van Speybroeck, "X-ray evidence for particulate contamination on the AXAF VETA-1 mirrors", Proc. SPIE, **1742**, 171–182, 1993.

[34] R. F. Elsner, S. L. O'Dell, & M. C. Weisskopf, "Molecular contamination and the calibration of AXAF", Proc. SPIE, **1742**, 6–12, 1993.

[35] D. E. Graessle, T. H. Burbine, J. J. Fitch, W. A. Podgorski, J. Z. Juda, R. F. Elsner, S. L. O'Dell, & J. M. Reynolds, "Molecular contamination study of iridium-coated x-ray mirrors", Proc. SPIE, **2279**, 12–26, 1994.

[36] K. L. Dietz, R. F. Elsner, M. K. Joy, S. L. O'Dell, B. D. Ramsey, M. C. Weisskopf, A. W. Armstrong, B. L. Colborn, & N. Kanvec, "Shielding imulations for the Advanced X-ray Astrophysics Facility (AXAF)", Proc. SPIE, **2518**, 107–118, 1995.

[37] R. F. Elsner, , M. K. Joy, S. L. O'Dell, B. D. Ramsey, & M. C. Weisskopf, "Ground-to-orbit transfer of the AXAF-I flux scale: In-situ contamination monitoring of x-ray telescopes", Proc. SPIE, **2279**, 332–342, 1994.

[38] R. F. Elsner, S. L. O'Dell, B. D. Ramsey, A. F. Tennant, M. C. Weisskopf, J. J. Kolodziejczak, D. A. Swartz, D. E. Engelhaupt, G. P. Garmire, J. A. Nousek, M. W. Bautz, T. J. Gaetz, & P. Zhao, "Calibration results for the AXAF flight contamination monitor", Proc. SPIE, **3444**, 177–188, 1998.

[39] R. F. Elsner, J. J. Kolodziejczak, S. L. O'Dell, D. A. Swartz, A. F. Tennant, & M. C. Weisskopf, "Measurements with the Chandra X-ray Observatory's flight contamination monitor", Proc. SPIE, **4012**, 612–618, 2000.

[40] R. F. Elsner, J. J. Kolodziejczak, S. L. O'Dell, D. A. Swartz, A. F. Tennant, & M. C. Weisskopf, "Measurements with the Chandra X-ray Observatory's flight contamination monitor", Proc. SPIE, **4138**, 1–9, 2000.

[41] M. C. Weisskopf & S. L. O'Dell, "Calibration of the AXAF observatory: Overview", Proc. SPIE, 3113, 2–17, 1997.

[42] S. L. O'Dell & M. C. Weisskopf, "Advanced X-ray Astrophysics Facility (AXAF): Calibration overview", Proc. SPIE, **3444**, 2–18, 1998.

[43] S. L. O'Dell, M. W. Bautz, W. C. Blackwell, Y. M. Butt, R. A. Cameron, R. F. Elsner, M. S. Gussenhoven, J. J. Kolodziejczak, J. I. Minow, R. M. Suggs, D. A. Swartz, A. F. Tennant, S. N. Virani, & K. M. Warren, "Radiation environment of the Chandra X-ray Observatory", Proc. SPIE, **4140**, 99–110, 2000.

[44] J. J. Kolodziejczak, R. F. Elsner, R. A. Austin, & S. L. O'Dell, "Ion transmission to the focal plane of the Chandra X-ray Observatory", Proc. SPIE, **4140**, 135–143, 2000.

[45] S. L. O'Dell, W. C. Blackwell, R. A. Cameron, J. I. Minow, D. C. Morris, B. J. Spitbart, D. A. Swartz, S. N. Virani, & S. J. Wolk, "Managing radiation degradation of CCDs on the Chandra X-ray Observatory", Proc. SPIE, **4851**, 77–88, 2003.

[46] S. L. O'Dell, T. L. Aldcroft, B. A. Bissell, W. C. Blackwell, R. A. Cameron, J. H. Chappell, J. M. DePasquale, K. R. Gage, C. E. Grant, C. F. Harbison, M. Juda, K. A. Marsh, E. R. Martin, J. I. Minow, S. S. Murray, P. P. Plucinsky, D. A. Schwartz, D. P. Shropshire, B. J. Spitzbart, S. N. Virani, B. S. Williams, & S. J. Wolk, "Managing radiation degradation of CCDs on the Chandra X-ray Observatory II", Proc. SPIE, **5898**, 212–223, 2005.

[47] W. C. Blackwell, J. I. Minow, S. L. O'Dell, R. M. Suggs, D. A. Swartz, A. F. Tennant, S. N. Virani, & K. M. Warren, "Modeling the Chandra space environment", Proc. SPIE, **4140**, 111–122, 2000.

[48] P. P. Plucinsky, S. L. O'Dell, N. W. Tice, D. A. Swartz, M. W. Bautz, J. M. DePasquale, R. J. Edgar, G. P. Garmire, R. Giordano, C. E. Grant, P. Knollenberg, S. Kissel, B. LaMarr, R. Logan, M. Mach, H. L. Marshall, L. McKendrick, G. Y. Prigozhin, D. Schwartz, N. S. Schulz, D. Shropshire, T. Trinh, A. A. Vikhlinin, & S. N. Virani, "An evaluation of a bake-out of the ACIS instrument on the Chandra X-ray Observatory", Proc. SPIE, **5488**, 251–263, 2004.

[49] S. L. O'Dell, D. A. Swartz, P. P. Plucinsky, M. A. Freeman, M. L. Markevitch, A. A. Vikhlinin, K. C. Chen, R. J. Giordano, P. J. Knollenberg, P. A. Morris, H. Tran, N. W. Tice, & S. K. Anderson, "Modeling contamination migration on the Chandra X-ray Observatory", Proc. SPIE, **5898**, 313–324, 2005.

[50] H. L. Marshall, A. Tennant, C. E. Grant, A. P. Hitchcock, S. L. O'Dell, & P. P. Plucinsky, "Composition of the Chandra ACIS contaminant", Proc. SPIE, **5165**, 497–508, 2004.

[51] A. F. Tennant, W. Becker, M. Juda, R. F. Elsner, J. J. Kolodziejczak, S. S. Murray, S. L. O'Dell, F. Paerels, D. A. Swartz, N. Shibazaki, & M. C. Weisskopf, "Discovery of X-Ray Emission from the Crab Pulsar at Pulse Minimum", Astrophys. J., **554**, L173–L176, 2001.